\documentclass[aps,prd,nofootinbib,floatfix,amsmath,amssymb,colorBG,spanish]{revtex4}
\usepackage{amssymb}
\usepackage{graphicx}
\usepackage{color}
\usepackage[tight,TABTOPCAP]{subfigure}
\usepackage{epstopdf}
\begin{document}

\title{Heavy neutral pseudoscalar decays into gauge bosons in the Littlest Higgs Model}

\author{$^{(a)}$J. I. Aranda, $^{(a)}$E. Cruz-Albaro, $^{(a)}$D. Espinosa-G\'omez, $^{(b,c)}$J. Monta\~no,
$^{(a)}$F. Ram\'irez-Zavaleta, $^{(a)}$E. S. Tututi}

\address{$^{(a)}$Facultad de Ciencias F\'isico Matem\'aticas, Universidad Michoacana de San Nicol\'as de Hidalgo,
Avenida Francisco J. M\'ujica S/N, 58060, Morelia, Michoac\'an, M\'exico\\
$^{(b)}$Instituto de F\'\i sica Te\'orica--Universidade Estadual Paulista, R. Dr. Bento Teobaldo Ferraz 271, Barra Funda, S\~ao Paulo - SP, 01140-070,
Brazil\\
$^{(c)}$C\'atedras Consejo Nacional de Ciencia y Tecnolog\'ia, M\'exico}

\date{\today}
\begin{abstract}
We study two-body decays of a new neutral pseudoscalar into gauge bosons within the context of the Littlest
Higgs model. Concretely, the $\Phi^P \to WW, VV, gg$ processes induced at the one-loop level, with $V=\gamma, Z$, are considered. Since the branching ratios of the $\Phi^P \to VV$ decays result very suppressed, only the $\Phi^P \to WW, gg$ processes are thoroughly studied. The branching ratios for the $\Phi^P \to gg$ and $\Phi^P \to WW$ decays are of the order of $10^{-4}$ and $10^{-6}$, respectively, for $f$ around 2 TeV, which represents the global symmetry breaking scale of the theory. The production cross section of the $\Phi^P$ boson via gluon fusion at LHC is estimated.

\end{abstract}

\pacs{12.60.-i, 14.70.-e, 14.80.Ec}

\maketitle
\section{Introduction}
After the discovery of the Higgs boson, the ATLAS~\cite{Aad:2012tfa} and CMS~\cite{Chatrchyan:2012xdj} collaborations have continued searching for new exotic particles, such as the Randall-Sundrum spin-2 boson or new heavy scalar particles~\cite{CMS:2015dxe, atlas1}. Particularly, ATLAS collaboration is carrying out searches for new scalar resonances decaying into two photons. Though up to the moment the searches for new scalar resonances has been fruitless, it is expected that the experimental collaborations ATLAS and CMS continue the seek for new particles at the TeV energy scale. These quests are supported by the improvements implemented in the CMS and ATLAS detectors~\cite{ATLAS:2013hta,Khachatryan:2014hpa,CMS:2014wda,Aad:2015owa} along with the fact that the experimental collaborations have been able to develop a reliable detection machinery of spin-0 resonances. Thus, the perspectives of searching for new physics phenomena at the TeVs scale are hopeful. Complementarily, several theoretical studies are found in the literature where the predictive power of extended models is tested via production of new-heavy neutral scalars decaying into standard model (SM) particles~\cite{MTS}.

The aforementioned experimental results offer theoretical arguments to explore new physics process related with extended Higgs sectors or SM extensions at the TeV energy scale~\cite{2HDM,LR1,LR2,MSSM1,Instantons,MSSM2,ZP,331M,3311M}. In this meaning, there exist many theoretical approaches that predict more content of scalar particles, such as the two-Higgs doublet models (2HDMs)~\cite{2HDM, 2HDMs1}, three-Higgs doublet model (3HDM)~\cite{3HDMs}, Higgs-singlet extension model~\cite{Pruna}, little Higgs models (LHMs)~\cite{LHMs}, etc. Among the variety of LHMs, the case of the littlest Higgs model (LTHM) is very interesting since it does no present new degrees of freedom beyond the SM under TeV scale. Additionally, the LTHM has a reduced spectrum of new scalar particles~\cite{LHM1}. As value added, this type of models offer a possible solution to hierarchy problem, which emerges when the Higgs boson mass is affected by one-loop corrections. Also, the LHMs represents another approach to the electroweak symmetry breaking pattern~\cite{LHMs,LHM1}, based on the dimensional deconstruction~\cite{Dec1,Dec2}, where the quadratic divergences cancel. Expressly, the quadratic divergence generated at the one-loop level because of the SM gauge bosons is canceled by the quadratic divergence introduced by the new gauge bosons at the same perturbative level. The one-loop quadratic divergence induced in the SM Yukawa sector is eliminated by introducing new heavy fermions in such a way that the quadratic divergence coming from the SM top quark cancels. In LHMs, in agreement with the breaking of the global symmetry at the energy scale of TeVs, the new Higgs field get mass becoming pseudo-Goldstone bosons, where in addition a massless Higgs arises. The quadratic-divergent corrections to the Higgs boson mass show up at loop level, which assures a light Higgs boson. With regard to LTHM~\cite{Han:2003wu}, the new particles arising at the TeV energy scale are grouped in a new set of four gauge bosons with the same quantum numbers as the SM gauge bosons, namely, $A_H$, $Z_H$, and $W^{\pm}_H$, an exotic quark with the same charge as the SM top quark, and a new heavy scalar triplet, which contains six physical states: double-charged scalars $\Phi^{\pm\pm}$, single-charged scalars $\Phi^\pm$, a neutral scalar $\Phi^0$, and a neutral pseudo scalar $\Phi^p$. All the details of the LTHM model can be found in Ref.~\cite{Han:2003wu}.

The aim of this work is to study in detail the one-loop level $\Phi^P\to WW, VV, gg$ ($V=\gamma, Z$) decays in the context of the LTHM. In particular, our calculations are placed on the basis of the so-called linearized theory of LTHM~\cite{Han:2003wu}, which means that it is performed a first-order expansion of the $\Sigma$ field around its vacuum expectation value (VEV) in powers of $v/f$~\cite{Han:2003wu}. Regarding the $\Phi^P WW$, $\Phi^P VV$ and $\Phi^P gg$ vertexes, at the one-loop level, these present a completely different Lorentz structure from what it has in similar interactions with a scalar particle~\cite{us}. Therefore, it is interesting to deepen the analytical study of the transition amplitudes for these interactions. Although the parameter space of the LTHM has been severely restricted by the Higgs discovery channels and electroweak precision observables~\cite{Reuter}, our proposal could be of interest as far as of the search for new scalar particles refers. Our study takes in consideration joined results from experimental and phenomenological analyzes, where it is proposed a simulated scenario as realistic as possible~\cite{Reuter}. The viable analysis region for the energy scale $f$ is sustained by a phenomenological study based on the experimental searches of the SM-Higgs boson, where it is used the signal strength modifier along with electroweak precision data. Even when the parameter space of the LTHM is strongly constrained, there is still room to study a wide variety of processes at the TeV energy scale. For consistency with electroweak precision data, a lower limit for the energy scale $f$ around 2-4 TeV is set~\cite{Reuter}.

The present paper is organized as follows. In section~\ref{MOD-FRA}, a survey of the LTHM is presented. In section~\ref{ANA-CAL}, the analytical calculations of the one-loop level $\Phi^P \to WW, VV, gg$ decays are described. In section~\ref{NUM-RES}, the numerical results are discussed. Finally, the conclusions are summarized in section~\ref{CONCLU}.


\section{Theoretical framework}\label{MOD-FRA}
The LTHM is established by a nonlinear sigma model with $SU(5)$ global symmetry together with the gauged group $[SU(2)_1\otimes U(1)_1]\otimes [SU(2)_2\otimes U(1)_2]$~\cite{LHM1, Han:2003wu}. At first, the $SU(5)$ group is spontaneously broken to the $SO(5)$ group at the energy scale $f$. At par, the $[SU(2)_1\otimes U(1)_1]\otimes [SU(2)_2\otimes U(1)_2]$ group is broken to its subgroup $SU_L(2)\otimes U_Y(1)$, the latter being the SM electroweak gauge group. At the energy scale $f$, the spontaneous global symmetry breaking of the $SU(5)$ group is generated by the VEV of the $\Sigma$ field, identified as $\Sigma_0$~\cite{Han:2003wu}. The explicit form of the $\Sigma$ field is given by
\begin{equation}
\Sigma=e^{i\Pi/f}\Sigma_0e^{i\Pi^T/f},
\end{equation}
with
\begin{equation}
\Sigma_0=\left(\begin{array}{ccc}
\mathbf{0}_{2\times 2} & \mathbf{0}_{2\times 1} & \mathbf{1}_{2\times 2}\\
\mathbf{0}_{1\times 2} & 1 & \mathbf{0}_{1\times 2}\\
\mathbf{1}_{2\times 2} & \mathbf{0}_{2\times 1} & \mathbf{0}_{2\times 2}
\end{array}\right)
\end{equation}
and $\Pi$ being the Goldstone boson matrix with the following structure
\begin{equation}
\Pi=\left(\begin{array}{ccc}
\mathbf{0}_{2\times 2} & h^\dagger/\sqrt{2} & \phi^\dagger\\
h/\sqrt{2} & 0 & h^\ast/\sqrt{2}\\
\phi & h^T/\sqrt{2} & \mathbf{0}_{2\times 2}
\end{array}\right).
\end{equation}
Here, the fields $h$ and $\phi$ represent a doublet and a triplet under the $SU_L(2)\otimes U_Y(1)$ SM gauge group, respectively~\cite{Han:2003wu}, having the following representation
\begin{equation}
h = (h^+, h^0), \qquad
\phi = \left( \begin{array}{cc}
\phi^{++} & \frac{\phi^+}{\sqrt{2}} \\
\frac{\phi^+}{\sqrt{2}} & \phi^0
\end{array} \right).
\end{equation}
The global symmetry breaking produces 14 Goldstone bosons which transform under the $SU_L(2)\otimes U_Y(1)$ group as a real singlet $\mathbf{1}_0$, a real triplet $\mathbf{3}_0$, a complex doublet $\mathbf{2}_{\pm \frac{1}{2}}$, and a complex triplet $\mathbf{3}_{\pm 1}$~\cite{LHM1,Han:2003wu}.
By means of this mechanism both the real singlet and the real triplet are absorbed by the longitudinal components of the gauge bosons at the scale $f$, where the complex doublet and the complex triplet remain massless. The complex doublet is precisely the SM Higgs field. Through the Coleman-Weinberg type potential when the global symmetry of the group $SO(5)$ breaks down the complex triplet gets mass of the order of the energy scale $f$.

On the other hand, the effective Lagrangian invariant under the $[SU(2)_1\otimes U(1)_1]\otimes [SU(2)_2\otimes U(1)_2]$
group is~\cite{Han:2003wu}
\begin{equation}
\mathcal{L}_{LTHM}=\mathcal{L}_G+\mathcal{L}_F+\mathcal{L}_\Sigma+\mathcal{L}_Y-V_{CW},
\end{equation}
where $\mathcal{L}_G$ contains the gauge bosons kinetic contributions, $\mathcal{L}_F$ represents the
fermions kinetic contributions, $\mathcal{L}_\Sigma$ includes the nonlinear sigma model contributions
of the LTHM, $\mathcal{L}_Y$ comprises the Yukawa couplings of fermions and pseudo-Goldstone bosons. The last term corresponds to the Coleman-Weinberg potential. For the purposes of our work, we only present a brief overview of the LTHM. For a more detailed description of this model see Ref.~\cite{Han:2003wu}.

The nonlinear sigma model sector is described by the following Lagrangian
\begin{equation}
\mathcal{L}_\Sigma=\frac{f^2}{8}\mathrm{tr}\left|\mathcal{D}_\mu\Sigma\right|^2,
\end{equation}
where the covariant derivative is expressed as
\begin{equation}
\mathcal{D}_\mu\Sigma=\partial_\mu\Sigma-i\sum\limits_{j=1}^{2}\left[g_j \sum\limits_{a=1}^{3}W_{\mu j}^{a}\left(Q_j^a\Sigma+\Sigma Q_j^{a T}\right)+g^\prime_j B_{\mu j}\left(Y_j\Sigma+\Sigma Y_j^{T}\right)\right].
\end{equation}
Explicitly, the $W_{\mu j}^{a}$ represent $SU(2)$ gauge fields, the $B_{\mu j}$ are the $U(1)$ gauge fields, $Q_j^a$ are the generators of the $SU(2)$ gauge group, the $Y_j$ denote $U(1)$ gauge group generators, $g_j$ are the coupling constants of the $SU(2)$ group, and $g_j^\prime$ are the coupling constants of the $U(1)$ group~\cite{Han:2003wu}. After the spontaneous symmetry breaking (SSB) around $\Sigma_0$, for the gauge bosons, mass eigenstates of order of $f$ are generated~\cite{Han:2003wu}
\begin{eqnarray}
W^\prime_\mu&=&-cW_{\mu 1}+sW_{\mu 2},\\
B^\prime_\mu&=&-c^\prime \mathcal{B}_{\mu 1}+s^\prime \mathcal{B}_{\mu 2},\\
W_\mu&=& sW_{\mu 1}+cW_{\mu 2},\\
B_\mu&=& s^\prime \mathcal{B}_{\mu 1}+c^\prime \mathcal{B}_{\mu 2}.
\end{eqnarray}
Here, $W_{\mu j}\equiv\sum\limits_{a=1}^{3}W^a_{\mu j}Q^a_j$ and $\mathcal{B}_{\mu j}\equiv B_{\mu j}Y_j$, for $j=1,2$; Moreover, $c=g_1/\sqrt{g_1^2+g_2^2}$, $c^\prime=g^\prime_1/\sqrt{g_1^{\prime 2}+g_2^{\prime 2}}$, $s=g_2/\sqrt{g_1^2+g_2^2}$, and $s^\prime=g^\prime_2/\sqrt{g_1^{\prime 2}+g_2^{\prime 2}}$. In the procedure of SSB, the $\Sigma$ field is expanded around its VEV ($\Sigma_0$) preserving dominant terms in the Lagrangian of the nonlinear sigma model sector~\cite{Han:2003wu}. Consequently, the masses of the new heavy gauge bosons are of order $\mathcal{O}\left(f\right)$
\begin{eqnarray}
m_{Z_H}&=&\frac{gf}{2sc},\\
m_{A_H}&=&\frac{g^\prime f}{2\sqrt{5}s^\prime c^\prime},\\
m_{W_H}&=&\frac{gf}{2sc}.
\end{eqnarray}
At this stage of the SSB the $B_\mu$ and $W_\mu$ fields still do not acquire mass. In the LTHM, the $c$ parameter, $c=m_{W_{H}}/m_{Z_{H}}$, takes the value closes to one at the leading order~\cite{Han:2003wu, Aranda}, in order to have similar values for the new gauge bosons masses; as it happens in the electroweak sector of the SM~\cite{Aranda}. While the $c^\prime$ parameter is related to the mixing angles between gauge bosons of the $SU(2)$ and $SU(1)$ groups, it directly affects the mass of the heavy photon ($A_H$). Therefore, in desire of keep masses of the order of TeVs, the $c^\prime$ parameter should be restricted to large mixing angles. At the Fermi energy scale, the SM gauge bosons get masses by the SSB mechanism, where also mixings between SM and new heavy gauge bosons are induced.

With regard to the scalar sector, in the LTHM the Higgs potential is induced via one-loop radiative corrections at the leading order. This potential comprises contributions coming from gauge boson and fermion loops. When the $\Sigma$ field is expanded around its VEV into the nonlinear sigma Lagrangian the Coleman-Weinberg potential is achieved~\cite{Coleman:1973jx}
\begin{equation}
 V_{CW}=\lambda_{\phi^2}f^2\mbox{Tr}(\phi^\dagger \phi)+
 i\lambda_{h\phi h}f(h\phi^\dagger h^T-h^* \phi^\dagger h^\dagger) -
 \mu^2hh^\dagger+\lambda_{h^4}(hh^\dagger)^2.
\end{equation}
The $\lambda$'s quantities are given as
\begin{eqnarray}
\lambda_{\phi^2} &=& \frac{a}{2} \left[ \frac{g^2}{s^2c^2}
+ \frac{g^{\prime 2}}{s^{\prime 2}c^{\prime 2}} \right]
+ 8 a^{\prime} \lambda_1^2,
\nonumber \\
\lambda_{h \phi h} &=& -\frac{a}{4}
\left[ g^2 \frac{(c^2-s^2)}{s^2c^2}
+ g^{\prime 2} \frac{(c^{\prime 2}-s^{\prime 2})}
{s^{\prime 2}c^{\prime 2}} \right]
+ 4 a^{\prime} \lambda_1^2,
\nonumber \\
\lambda_{h^4} &=& \frac{a}{8} \left[ \frac{g^2}{s^2c^2}
+ \frac{g^{\prime 2}}{s^{\prime 2}c^{\prime 2}} \right]
+ 2 a^{\prime} \lambda_1^2 = \frac{1}{4} \lambda_{\phi^2}.
\end{eqnarray}
The parameters: $c$, $s$ ($c^\prime$, $s^\prime$) represent mixing angles related with the gauge coupling constants of the $SU(2)$ ($U(1)$) symmetry group. The $a$ and $a^\prime$ parameters depict unknown ultraviolet (UV) physics at the cutoff scale $\Lambda_S$. Their values depend on the UV completion details at the $\Lambda_S$ scale~\cite{Han:2003wu}. The $\mu$ quantity is a free parameter that receives evenly significant contributions coming from one-loop logarithmic and two-loop quadratically divergent parts~\cite{Han:2003wu}. The VEV $v$ ($v^\prime$) of the doublet (of the triplet) is obtained after minimizing the Coleman-Weinberg potential, where there is fulfilled the relations
\begin{equation}
v^2=\frac{\mu^2}{\lambda_{h^4} -\frac{\lambda^2_{h\phi h}}{\lambda_{\phi^2}}},
\qquad v^\prime=\frac{\lambda_{h\phi h}v^2}{2\lambda_{\phi^2} f}.
\end{equation}
By diagonalizing the Higgs mass matrix heavy scalar bosons get masses~\cite{Han:2003wu}. Thereby, the gauge eigenstates of the Higgs sector can be expressed depending on mass eigenstates in the following manner
\begin{eqnarray}
	h^0 &=& \frac{1}{\sqrt{2}}\left( c_0 H - s_0 \Phi^0 + v \right)
	+ \frac{i}{\sqrt{2}} \left( c_P G^0 - s_P \Phi^P \right), \nonumber \\
	\phi^0 &=& \frac{1}{\sqrt{2}}\left( s_P G^0 + c_P \Phi^P \right)
	- \frac{i}{\sqrt{2}} \left( s_0 H + c_0 \Phi^0 + \sqrt{2} v^{\prime} \right),
	\nonumber \\
	h^+ &=& c_+ G^+ - s_+ \Phi^+, \nonumber\\
	\phi^+ &=&\frac{1}{i}\left( s_+ G^+ + c_+ \Phi^+ \right),  \nonumber \\
	\phi^{++} &=& \frac{\Phi^{++}}{i}.
\end{eqnarray}
Here, $H$ symbolizes the Higgs boson, $\Phi^0$ is a new neutral scalar, $\Phi^P$ represents the neutral pseudoscalar, $\Phi^+$ and $\Phi^{++}$ are the charged and doubly charged scalars, respectively. The $G^+$ and $G^0$ fields are the Goldstone bosons that are eaten by the massless $W$ and $Z$ bosons~\cite{Han:2003wu}. At the leading order, the masses of the new scalar bosons are degenerate, being written as~\cite{Han:2003wu}
\begin{equation}
 m_\Phi=\frac{\sqrt{2} m_H}{\sqrt{{1-y_v^2}}} \frac{f}{v},
\label{massphi}
\end{equation}
where $y_v=4v^{\prime} f / v^2$. The previous expression is only definite positive if $\frac{v'^2}{v^2} < \frac{ v^2}{16f^2}$.

For the rest of particle content, the LTHM predicts new fermions which couple to the Higgs field in such a manner that the quadratic divergence of the top quark is annulled~\cite{LHM1,Han:2003wu}. The new set of heavy fermions is settled as a vector-like pair ($\tilde{t},\tilde{t}^{\prime c}$) with quantum
numbers $(\mathbf{3},\mathbf{1})_{Y_i}$ and $(\bar{\mathbf{3}},\mathbf{1})_{-Y_i}$, respectively. Thus, the Yukawa sector has the following structure
\begin{equation}
\mathcal{L}_Y=\frac{1}{2}\lambda_1\, f\,\epsilon_{ijk}\epsilon_{xy}\,\chi_i\,\Sigma_{jx}\,\Sigma_{ky}u_3^{\prime c}+\lambda_2\,f\,\tilde{t}\tilde{t}^{\prime c}+\mathrm{H.c.},
\end{equation}
where $\chi_i=(b_3,t_3,\tilde{t})$; $\epsilon_{ijk}$ and $\epsilon_{xy}$ are antisymmetric
tensors for $i,j,k=1,2,3$ and $x,y=4,5$~\cite{LHM1}. Here, $\lambda_1$ and $\lambda_2$ represent free
parameters, where $\lambda_2$ can be fixed in such a way that for given values of $(f,\lambda_1)$, the
top quark mass should adjust to its experimental measurement~\cite{Reuter}. By expanding the $\Sigma$ field around $\Sigma_0$ keeping terms up to $\mathcal{O}(v^2/f^2)$ and then diagonalizing the mass matrix, it can be found the mass states $t_L$, $t^c_R$, $T_L$, and $T^c_R$, which correspond to the SM top quark and the new top quark, respectively~\cite{Han:2003wu,Reuter}. All the remaining contributions to the LTHM Lagrangian can be consulted in Ref.~\cite{Han:2003wu}. Our analytical results were computed by using the set of new Feynman rules presented in Ref.~\cite{Han:2003wu}.

\section{The $\Phi^P\to WW, \gamma V, gg$ decays}\label{ANA-CAL}
In order to analyze the one-loop level decays of the $\Phi^P$ boson, the total decay width of the $\Phi^P$ ($\Gamma_{\Phi^P}$), which will include only SM final states, must be computed. The most important contributions to $\Gamma_{\Phi^P}$ are the tree-level decays of $\Phi^P$ into $ZH,WWZ$. Other subdominant contributions turn out to be the final states: $\bar{t}t,WWH$, which we will also consider.

\subsubsection{Tree-level decays of the $\Phi^P$ boson}
\begin{figure}[!ht]
\begin{center}
\includegraphics[scale=0.65]{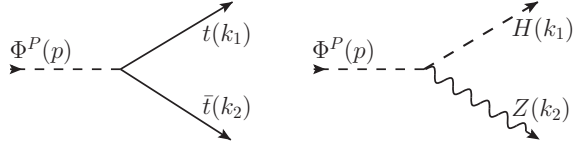}
\caption{Feynman diagrams corresponding to the $\Phi^P \to t \bar t, H Z$ decays at tree level.}
\label{feyn2VV}
\end{center}
\end{figure}

\begin{figure}[!ht]
\begin{center}
\includegraphics[scale=0.75]{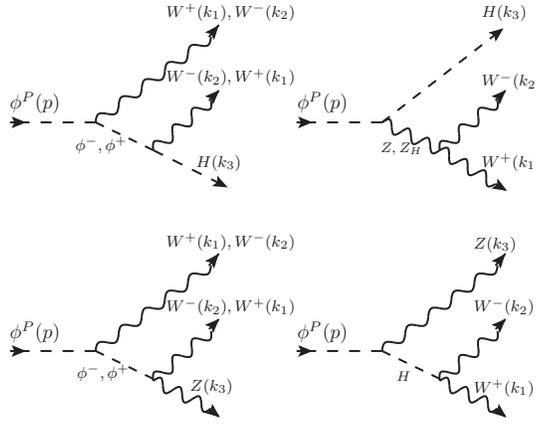}
\caption{Feynman diagrams representing three-body decays of the $\Phi^P$ boson at tree level.}
\label{feyn2ABC}
\end{center}
\end{figure}

The Feynman diagrams that depict two- and three-body decays of $\Phi^P$ boson at three level are shown in Figs.~\ref{feyn2VV} and \ref{feyn2ABC}, respectively.
On this basis, the decay width of the $\Phi^P \to t\bar t$ process is calculated, resulting in
\begin{equation}
\Gamma(\Phi^P \to \bar t t )= \frac{N_c\, m_{\Phi^P}\,m_t^2}{64\pi f^2}  \bigg(1-4\frac{m_t^2}{ m_{\Phi^P}^2}\bigg)^{1/2},
\end{equation}
where $m_{\Phi^P}$ is the mass of the pseudoscalar boson, $N_c$ is the color factor equal to 3 for quarks and $f$ is the global symmetry breaking scale.

The $\Phi^P\to ZH$ decay width can be written as follows
\begin{eqnarray}
\Gamma(\Phi^P \to ZH)&=& \frac{g^2v^2 m_{\Phi^P}^3}{512\,\pi\, c_W^2\,m_{Z}^2 f^2} \Bigg[ \bigg(1-\bigg(\frac{m_H-m_Z}{m_{\Phi^P}}\bigg)^2\bigg)\bigg(1-\bigg(\frac{m_H+m_Z}{m_{\Phi^P}}\bigg)^2\bigg)\Bigg]^{3/2}.
\end{eqnarray}

Concerning the three-body decays, these processes are mediated by SM particles, new scalar bosons and the $Z_H$ gauge boson (see Fig.~\ref{feyn2ABC}). The analytic expression for the decay width of the $\Phi^P$ scalar boson decaying into three bodies can be computed by using the standard formulation described in Ref.~\cite{PDG}.

The amplitude for the $\Phi^P \to WWH$ decay is
\begin{eqnarray}
\mathcal{M}(\Phi^P \to WWH)&=&\frac{g^2v}{8\sqrt{2}f}\Bigg[\frac{(k_1^{\mu}+2(k_2^{\mu}+k_3^{\mu}))(k_2^{\nu}+2k_3^{\nu})}{(k_2+k_3)^2-m_{\Phi^P}^2}
-\frac{(2k_1^{\nu}+k_2^{\nu}+2k_3^{\nu})(k_1^{\mu}+2k_3^{\mu})}{(k_1+k_3)^2-m_{\Phi^P}^2} -2i\left(k_1^{\alpha}+k_2^{\alpha}+2 k_3^{\alpha}\right)\nonumber\\
&\times&\left(\frac{-g^{\alpha\beta}+\frac{(k_1^{\alpha}+k_2^{\alpha})(k_1^{\beta}+k_2^{\beta})}{m_Z^2}}{(k_1+k_2)^2-m_Z^2}\right)
\bigg((2k_1^{\nu}+k_2^{\nu})g^{\beta\mu}
+(k_2^{\beta}-k_1^{\beta})g^{\nu\mu}-(k_1^{\mu}+2k_2^{\mu})g^{\beta\nu}\bigg)\nonumber\\
&-&i\frac{(c^2-s^2)(c_W x^{W'}_Z + sc(s^2-c^2))v^2}{csf^2}\Bigg\{\big(k_1^{\alpha}+k_2^{\alpha}
+2k_3^{\alpha}\big)\left(\frac{-g^{\alpha\beta}+\frac{(k_1^{\alpha}+k_2^{\alpha})(k_1^{\beta}+k_2^{\beta})}{m^2_{Z_H}}}{(k_1+k_2)^2-m^2_{Z_H}}\right)\nonumber\\
&\times&\bigg((2k_1^{\nu}+k_2^{\nu})g^{\beta\mu}
+(k_2^{\beta}-k_1^{\beta})g^{\nu\mu}-(k_1^{\mu}+2k_2^{\mu})g^{\beta\nu}\bigg) \Bigg\}   \Bigg] \epsilon^{*}_{\mu}(k_1)\epsilon^{*}_{\nu}(k_2),
\end{eqnarray}
where $x^{W'}_Z=-\frac{1}{2c_W} sc(c^2-s^2)$. Due to degeneracy in the masses of new scalar bosons, $m_{\Phi^P}=m_{\Phi^+}=m_{\Phi^-}$ was employed.
\begin{eqnarray}
\mathcal{M}(\Phi^P \to WWZ)&=&\frac{g^3v^2}{8\sqrt{2}c_Wf}\Bigg[\frac{\bigg(2k_1^{\nu}+k_2^{\nu}+2k_3^{\nu}\bigg)g^{\mu\alpha}}{(k_1+k_3)^2-m_{\Phi^P}^2}
-\frac{\bigg(k_1^{\mu}+2(k_2^{\mu}+k_3^{\mu})\bigg)g^{\nu\alpha}}{(k_2+k_3)^2-m_{\Phi^P}^2}-\frac{\bigg(2(k_1^{\alpha}+k_2^{\alpha})+k_3^{\alpha}\bigg)g^{\mu\nu}}{(k_1+k_2)^2-m_H^2} \Bigg]\nonumber\\
&\times&\epsilon^{*}_{\mu}(k_1)\epsilon^{*}_{\nu}(k_2)\epsilon^{*}_{\alpha}(k_3).
\end{eqnarray}
Owing to the fact that the main impact of new particles on the different decay modes of the $\Phi^P$ boson would be manifesting at tree level, we will analyze this type of contributions exclusively at this level. For this purpose, it is necessary to fix the $c$ parameter or at least to explore a neighborhood for it with the objective of maximizing the possible contributions of new physics. Particularly, our election seeks to avoid the decoupling in the $\Phi^P Z_H H$ interaction. Thereby, it is assumed that $c\approx0.85$~\cite{us}, which indicates a small mixing angle and promotes an enhanced contribution of the $Z_H$ boson to the $\Phi^P\to WWH$ decay. The numerical results tell us that new heavy particle corrections are less than 1\%. All the analytical formulae necessary to calculate the decay widths were computed by using the FeynCalc package~\cite{FeyC}.

\subsubsection{One-loop level decays of the $\Phi^P$ boson}
Firstly, we will discuss relevant details of $\Phi^P WW$ vertex in order to present the analytic results for the $\Phi^P\to WW$ decay. Secondly, the analytical study of the $\Phi^P VV$ vertexes, where ($V=\gamma, Z$), will be placed in the context of $\Phi^P\to VV$ decays. Lastly, analytical results for the one-loop amplitude of the $\Phi^P\to gg$ process are presented.

\begin{figure}[!ht]
\begin{center}
\includegraphics[scale=0.50]{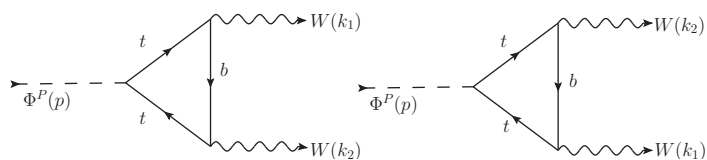}
\caption{Feynman diagrams contributing to the $\Phi^P \to WW$ process at one-loop level.}
\label{diaWW}
\end{center}
\end{figure}

Motivated by the fact that the $\Phi^P WW$ vertex is absent at tree-level in the LTHM, it is interesting to analyze this coupling at one-loop level. This can be done through the analysis of the $\Phi^P \to WW$ process. In what follows, all the one-loop calculations will be carried out by making use of the unitary gauge. In the LTHM, this decay only receives contributions from quarks, however, the main one corresponds to SM top quark. The dominant Feynman diagrams that represent the  $\Phi^P \to WW$ decay are shown in Fig.~\ref{diaWW}. After performing dimensional regularization for the one-loop amplitudes related with the Feynman diagrams in Fig.~\ref{diaWW}, we find that the total amplitude for the $\Phi^P \to WW$ decay can be written as
\begin{eqnarray}\label{PhiWW}
\mathcal{M}(\Phi^P \to WW)&=&A^{WW} \epsilon^{\mu \nu\alpha\beta}{k_1}_\alpha {k_2}_\beta \epsilon^{*}_{\mu}(k_1)\epsilon^{*}_{\nu}(k_2),
\end{eqnarray}
where
\begin{eqnarray}
A^{WW}&=&\frac{g^2\,N_{C}\,m_t^2\vert V_{tb} \vert^2}{8\sqrt{2}\,\pi^2\,f(4m_W^2-m_{\Phi^P}^2)} \bigg[\big(B_0(1)-B_0(2)\big)
+\big(m_b^2-m_t^2+m_W^2\big) C_{0}(1)\bigg].
\end{eqnarray}
It should be stressed that the form factor $A^{WW}$ is finite, being, $B_0(1)=B_0(m_{\Phi^P}^2,m_t^2,m_t^2), B_0(2)=B_0(m_{W}^2,m_b^2,m_t^2)$ and $C_0(1)=C_{0}(m_W^2,m_W^2,m_{\Phi^P}^2,m_t^{2},m_b^{2},m_t^{2})$, the Passarino-Veltman scalar functions (PV). Thus, the decay width of the $\Phi^P \to WW$ process is
\begin{equation}\label{dwidthWW}
\Gamma(\Phi^P \to WW)=\frac{1}{32\,\pi} \left|A^{WW}\right|^2 (m_{\Phi^P}^2-4m_W^2)^{\frac{3}{2}}.
\end{equation}

\begin{figure}[!ht]
\begin{center}
\includegraphics[scale=0.45]{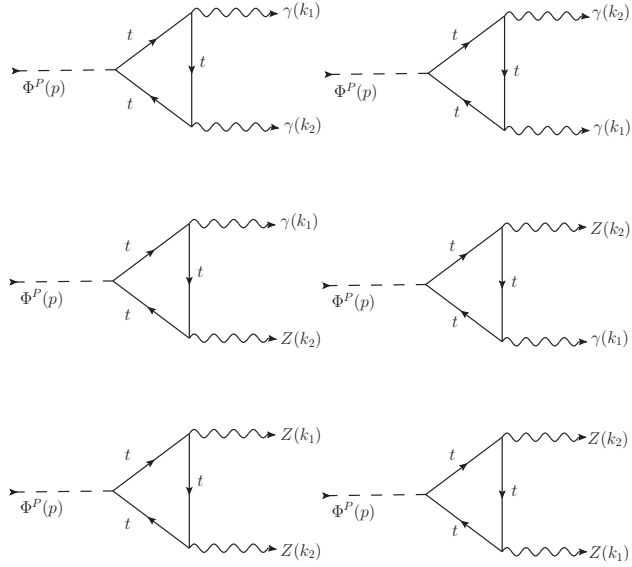}
\caption{Dominant Feynman diagrams that contribute to the $\Phi^P \to VV$ ($V=\gamma, Z$) decays at one-loop level.}
\label{unchargedweaks}
\end{center}
\end{figure}

The following describes the analytical expressions for the amplitudes of the $\Phi^P \to V V$ processes. In the LTHM context, the $\Phi^P \to V V$ decays are only mediated by SM quarks and the new exotic top quark $T$. Even though the new top quark induces a contribution different from zero, this is suppressed at least by two orders of magnitude with respect to the SM top quark contribution, which is the dominant one regarding the remaining SM quarks. Moreover, when one $T$ quark fluctuates into the $\Phi^P \to \gamma \gamma$ decay, its amplitude is exactly zero since there is no $Tt\gamma$ vertex. Therefore, we do not present this in the Feynman diagrams. In Fig. \ref{unchargedweaks} the Feynman diagrams associated with the $\Phi^P \to VV$ decays are shown.

The one-loop amplitude for the $\Phi^P \to \gamma \gamma$ decay can be written as follows
\begin{eqnarray}\label{phigaga}
\mathcal{M}(\Phi^P \to \gamma \gamma)=A^{\gamma \gamma} \epsilon^{\mu \nu \alpha \beta}{k_1}_\alpha {k_2}_\beta \epsilon^{*}_{\mu}(k_1)\epsilon^{*}_{\nu}(k_2).
\end{eqnarray}
The form factor $A^{\gamma \gamma}$ is given in terms of PV, which can be seen below
\begin{eqnarray}
A^{\gamma \gamma}=\frac{-g^2\,N_{C}\,s_W^2}{9\sqrt{2}\,\pi^2\,f}\,m_t^2 C_0(2),
\end{eqnarray}
where $C_0(2)=C_{0}(0,0,m_{\Phi^P}^2,m_t^{2},m_t^{2},m_t^{2})$ is the three-point PV. After performing algebraic operations, the decay width of the $\Phi^P \to \gamma \gamma$ decay is expressed as
\begin{eqnarray}
\Gamma(\Phi^P \to \gamma \gamma)=\frac{1}{64\pi} \left|A^{\gamma\gamma}\right|^2 m_{\Phi^P}^3.
\end{eqnarray}

In this way, the one-loop amplitude for the $\Phi^P\to \gamma Z$ decay can be computed, being equal to
\begin{eqnarray}\label{phigaz}
\mathcal{M}(\Phi^P \to \gamma Z)&=&A^{\gamma Z} \epsilon^{\mu \nu \alpha \beta}{k_1}_\alpha {k_2}_\beta\epsilon^{*}_{\mu}(k_1)\epsilon^{*}_{\nu}(k_2).
\end{eqnarray}
The form factor $A^{\gamma Z}$ has the following structure
\begin{eqnarray}
A^{\gamma Z}&=& \frac{g^2\,N_{C}\,s_W\,(3-8\,s_W^2)}{72\sqrt{2}\,\pi^2\,c_W\,f}\, m_t^2 C_0(3),
\end{eqnarray}
where $C_0(3)=C_{0}(m_Z^2,m_{\Phi^P}^2,0,m_t^{2},m_t^{2},m_t^{2})$. Thereby, the associated decay width can be expressed as follows
\begin{eqnarray}
\Gamma(\Phi^P \to \gamma Z)= \frac{1}{32\,\pi\,m_{\Phi^P}^3} \left|A^{\gamma Z}\right|^2 (m_{\Phi^P}^2-m_Z^2)^3.
\end{eqnarray}

The respective decay amplitude for the $\Phi^P \to Z Z$ process turns out to be
\begin{eqnarray}\label{phizz}
\mathcal{M}(\Phi^P \to Z Z)&=&A^{Z Z} \epsilon^{\mu \nu \alpha \beta}{k_1}_\alpha {k_2}_\beta\epsilon^{*}_{\mu}(k_1)\epsilon^{*}_{\nu}(k_2).
\end{eqnarray}
In this case, the form factor $A^{Z Z}$ is given by
\begin{eqnarray}
A^{Z Z}&=& \frac{g^2\,N_{C}\,m_t^2}{144\sqrt{2}\,\pi^2\,c_W^2(4m_Z^2-m_{\Phi^P}^2)\,f} \bigg[9\bigg(B_0(1)-B_0(3)\bigg)
+\bigg(4(3-4s_W^2)s_W^2\,m_{\Phi^P}^2+(3-8s_W^2)^2\,m_Z^2\bigg) C_{0}(4)\bigg],
\end{eqnarray}
where $B_0(3)=B_0(m_{Z}^2,m_t^2,m_t^2)$ and $C_0(4)=C_{0}(m_Z^2,m_Z^2,m_{\Phi^P}^2,m_t^{2},m_t^{2},m_t^{2})$. Following a similar algebraic procedure as above, the decay width of the $\Phi^P \to Z Z$ process can be obtained, being
\begin{eqnarray}
\Gamma(\Phi^P \to Z Z)= \frac{1}{64\,\pi} \left|A^{ZZ}\right|^2 (m_{\Phi^P}^2-4m_Z^2)^{\frac{3}{2}}.
\end{eqnarray}
It should be emphasized that the form factors: $A^{\gamma\gamma}, A^{\gamma Z}$ and $A^{Z Z}$ are free of ultraviolet (UV) divergences and its corresponding Lorentz structures satisfy gauge invariance.
\begin{figure}[!ht]
\begin{center}
\includegraphics[scale=0.5]{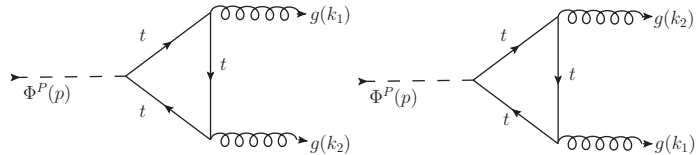}
\caption{Dominant Feynman diagrams that contribute to the $\Phi^P \to gg$ decay at one-loop level.}
\label{gluons}
\end{center}
\end{figure}

Finally, we exhibit the analytical result for the one-loop amplitude of the $\Phi^P \to gg$ decay. Also, for this process, we have only included the SM top quark contribution since the exotic top quark contribution is suppressed at least two orders of magnitude (see Fig.~\ref{gluons}). Said amplitude can be appreciated below
\begin{eqnarray}\label{phigg}
\mathcal{M}(\Phi^P \to gg)&=&A^{gg} \epsilon^{\mu \nu \alpha \beta}k_{1\alpha} k_{2\beta} \epsilon^{*a}_{\mu}(k_1)\epsilon^{*b}_{\nu}(k_2) \delta_{ab}.
\end{eqnarray}
The form factor $A^{gg}$ is written as follows
\begin{eqnarray}
A^{gg}&=& \frac{g_s^2}{8\sqrt{2}\,\pi^2\,f}\,m_t^2 C_0(5),
\end{eqnarray}
where $C_0(5)=C_{0}(m_{\Phi^P}^2,0,0,m_t^{2},m_t^{2},m_t^{2})$. In this manner, the decay width of the $\Phi^P \to gg$ process is given as
\begin{eqnarray}
\Gamma(\Phi^P \to gg)=\frac{1}{8\pi} \left|A^{gg}\right|^2  m_{\Phi^P}^3.
\end{eqnarray}
Notice that the form factor $A^{gg}$ is also UV finite and the respective amplitude complies gauge invariance.

The appearance of the Levi-Civita tensor in all the one-loop decay amplitudes of the $\Phi^P$ boson is a distinctive manifestation of the pseudoscalar nature of this particle~\cite{Kniehl}.

In all our analytical calculations it has been assumed that $v^\prime=\frac{v^2}{8f}$~\cite{us,Aranda}. The numerical evaluation for all the $\Phi^P \to WW, VV, gg$ processes was carried out by using the LoopTools package~\cite{LoopTools}.

\section{Numerical discussion}\label{NUM-RES}
This work represents a follow-up study to the one made by some of us concerning the two-body decays of the $\Phi^0$ scalar boson at one-loop level~\cite{us}, because now we are interested in investigating phenomenological details regarding the two-body decays of the pseudoscalar boson at one-loop level in the context of the LTHM. In this sense, a scenario where $m_{\Phi^P}$ is of the order of unities of TeVs is settled down, where the one-loop level decays of $\Phi^P$ into $WW, VV$ and $gg$ will be useful to test the consistency of the current parameter space of the LTHM~\cite{Reuter}. Before proceeding, we remark that now the energy scale $f$ is the only free parameter which we can play with. This symmetry-breaking energy scale is restricted by the experimental data to lower limits by around 2-4 TeV~\cite{Reuter}.

\begin{figure}[htb!]
\begin{center}
\includegraphics[scale=0.75]{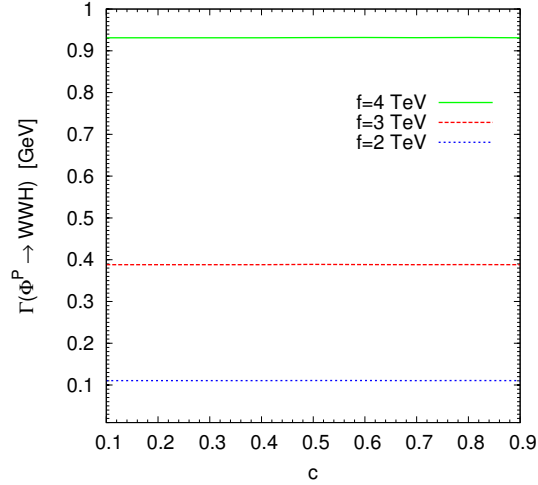}
\caption{Decay width of the $\Phi^P\to WWH$ process as a function of the $c$ parameter for $f=2,3,4$ TeV.}
\label{widthc}
\end{center}
\end{figure}

In Fig.~\ref{widthc}, it can be seen that the decay width of the $\Phi^P\to WWH$ process essentially does not depend on $c$ parameter variations. This plot shows a variation of the $c$ parameter from 0.1 to 0.9, for three distinct energy scales, i.e. $f=2$ TeV, $f=3$ TeV and $f=4$ TeV. Because of the $\Phi^P\to WWH$ decay is the only one that depends on the $c$ parameter, and this is not the dominant decay mode of the $\Phi^P$ boson, we assume that the total decay width of the $\Phi^P$ boson will not depend of the $c$ parameter. However, we will adopt a specific value for the $c$ parameter such that the $Z_H$ contribution to $\Phi^P\to WWH$ is enhanced. Thus, we fix $c=0.85$, in order to avoid the decoupling in the $\Phi^P Z_H H$ and $Z_H WW$ interactions, which also indicates presence of small mixing angles~\cite{us}.

\begin{figure}[htb!]
\begin{flushleft}
\subfigure[]{\includegraphics[width=8.6cm,height=7.0cm]{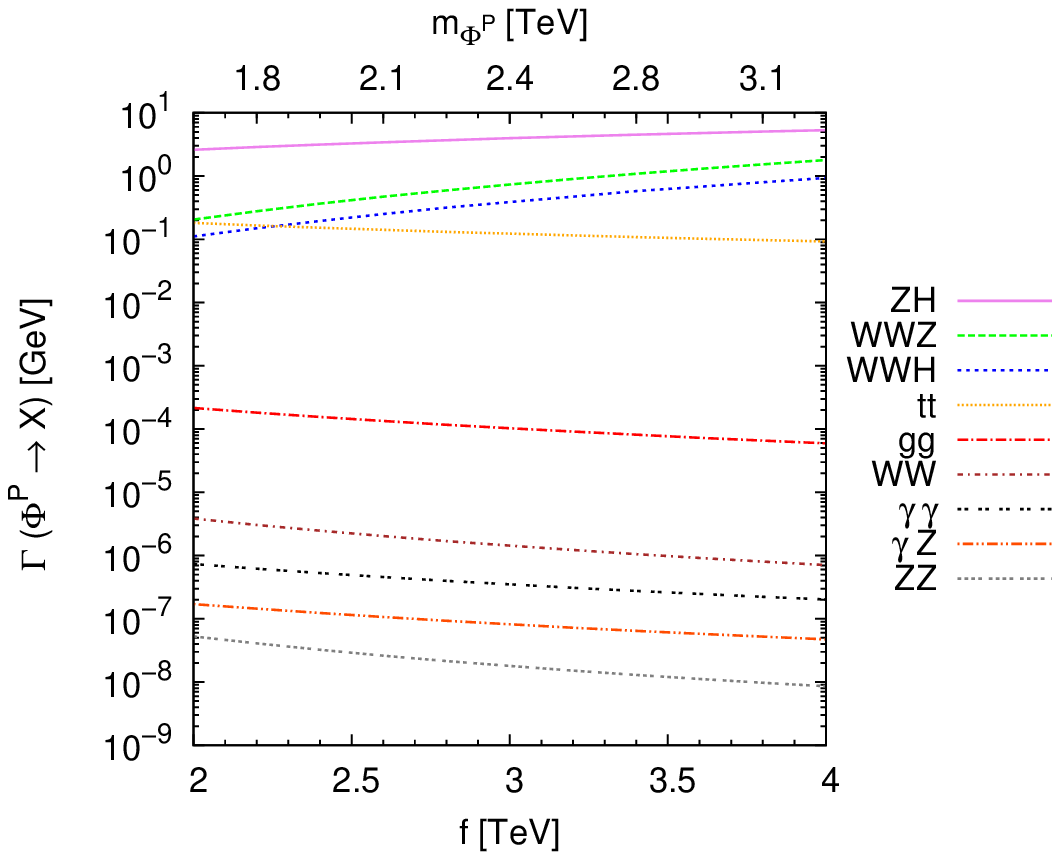}}\qquad
\subfigure[]{\includegraphics[width=8.6cm,height=7.0cm]{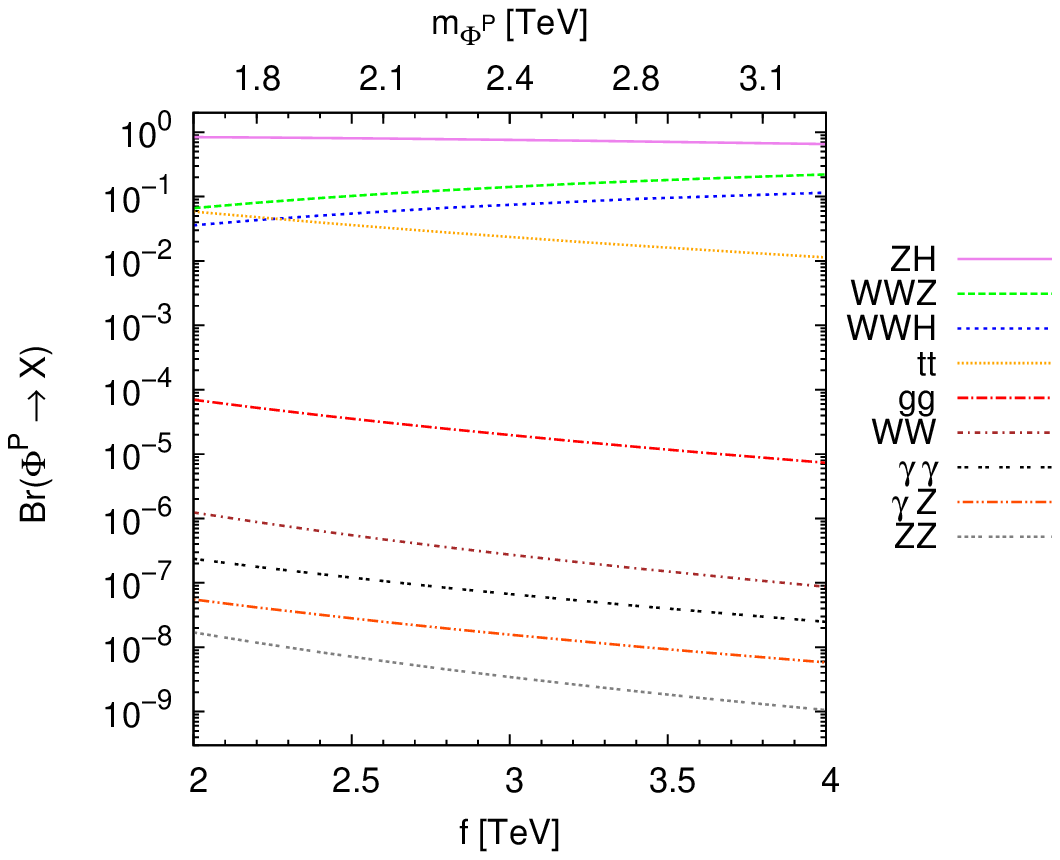}}
\caption{(a) Decay widths for the $\Phi^P\to X$ processes as a function of the $f$ energy scale, where $X=ZH, WWZ, WWH, tt, gg, WW, VV$. (b) Branching ratios for the same decays depending on the $f$ energy scale.}
\label{Brf}
\end{flushleft}
\end{figure}

In Fig.~\ref{Brf}(a), the decay widths for the $\Phi^P\to ZH, WWZ, WWH, tt, gg, WW, VV$ processes are displayed. From Fig.~\ref{Brf}(a), it can be observed that the dominant contributions come from tree-level decays of $\Phi^P$, specifically, into $ZH$, $WWZ$ and $WWH$, which are close to the value of 1 GeV around $f=4$ TeV. The two-body decays of the $\Phi^P$ boson at tree level have already been calculated~\cite{Twobwidhts}, however, between the energies studied only a small range is consistent with current constraints that come from electroweak precision data~\cite{Reuter}. At the one-loop level, we found a subdominant decay mode which corresponds to the $\Phi^P\to gg$ decay, whose associated decay width is of the order of $10^{-4}$ GeV near to $f=2$ TeV. The decay width of the $\Phi^P\to WW$ process is two orders of magnitude less than $\Gamma(\Phi^P\to gg)$, being of the order of $10^{-6}$ GeV close to $f=2$ TeV. In contrast, the numerical evaluation tell us that the decays to electroweak neutral bosons are very suppressed. In specific, $\Gamma(\Phi^P\to \gamma\gamma)\sim10^{-7}$ GeV, $\Gamma(\Phi^P\to \gamma Z)\sim10^{-7}$ GeV and $\Gamma(\Phi^P\to ZZ)\sim10^{-8}$ GeV for $f$ around 2 TeV. Furthermore, we have computed their corresponding branching ratios as a function of the energy scale $f$, as can be seen in Fig.~\ref{Brf}(b). The total decay width of the $\Phi^P$ pseudoscalar boson contains the decay modes: $ZH, WWZ, WWH, tt$. In Fig.~\ref{Brf}(b), it is clearly appreciated that the dominant branching ratios correspond to tree-level decays of $\Phi^P$. Nevertheless, the one-loop decay of the $\Phi^P\to gg$ process offers a branching ratio on the order of $10^{-4}$ for $f=2$ TeV. The remaining one-loop branching ratios turn out to be $\mathrm{Br}(\Phi^P\to WW)\sim10^{-6}$, $\mathrm{Br}(\Phi^P\to \gamma\gamma)\sim10^{-7}$, $\mathrm{Br}(\Phi^P\to \gamma Z)\sim10^{-7}$ and $\mathrm{Br}(\Phi^P\to ZZ)\sim10^{-8}$, for $f=2$ TeV. According to all the one-loop electroweak decay modes, the $\Phi^P\to WW$ process stands out as the most interesting one in order to search for new scalar bosons of pseudoscalar nature. However, in phenomenological terms, this is not the most important decay mode at one-loop level, since the $\Phi^P$ decay to two gluons would be the most likely to be found.

\subsection{An estimation for the production of the pseudoscalar boson}
In this section we present an approximate study for the production cross section of the exotic pseudoscalar $\Phi^P$ in the context fo the LTHM at LHC, decaying into different final states mentioned above. Although this analysis is not entirely accurate, it can provide valuable information as for the order of magnitude of the production cross section of the $\Phi^P$ particle is concerned. To perform this, we employ the Breit-Wigner resonant cross section~\cite{PDG}. In this approximation, the production cross section via gluon fusion can be computed by means of the branching ratios $\mathrm{Br}(\Phi^P\to gg)$ and $\mathrm{Br}(\Phi^P\to Y)$, where $Y=gg, WW, VV$. Thus, our Breit-Wigner cross section is written as follows
\begin{eqnarray}
\sigma(gg\to \Phi^P \to Y)=\frac{\pi}{12} \frac{Br(\Phi^P \to gg) Br(\Phi^P \to Y)}{m_{\Phi^P}^2}\label{n3},
\end{eqnarray}
where $\sigma(gg\to \Phi^P \to Y)$ is estimated just at the resonance of the pseudoscalar boson $\Phi^P$~\cite{PDG}.
\begin{figure}[htb!]
\includegraphics[scale=0.85]{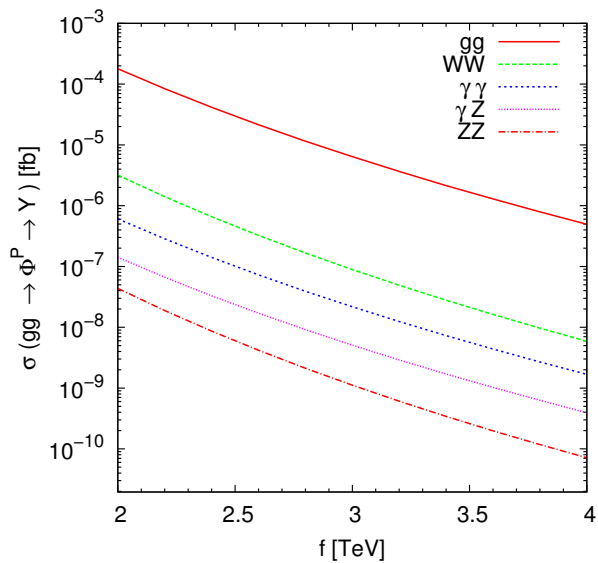}
\caption{Production cross section for $\Phi^P$ in gluon fusion as a function of the $f$ energy scale.
\label{phiprod}}
\end{figure}

We have numerically calculated the production cross section of $\Phi^P$ as a function of the $f$ parameter from $2$ TeV to $4$ TeV (see Fig.~\ref{phiprod}). From Fig.~\ref{phiprod}, it is evident that the region that has the greater predictive importance corresponds to $f$ around 2 TeV. Because the branching ratios for the $\Phi^P\to VV$ processes are very suppressed, we will focus our analysis on the decays $\Phi^P\to WW$ and $\Phi^P\to gg$. For $f=2$ TeV, the production cross section of $\Phi^P$ with $WW$ final states is $\sigma(gg\to \Phi^P \to WW)=3.18\times10^{-6}$ fb, whereas that $\sigma(gg\to \Phi^P \to gg)=1.78\times10^{-4}$ fb. The expected integrated luminosity of the LHC at the last stage of operation is projected to be around 3000 fb$^{-1}$~\cite{luminosity}. Therefore, by considering this experimental scenario it would be very difficult to observe some event related with the $\Phi^P\to WW$, nonetheless, there would still be an observation gap for the $\Phi^P \to gg$ process, since almost one event could be observed at LHC. If we compare our results with those obtained in the context of the $\Phi^0$ production at the LHC~\cite{us}, we find that the branching ratio for the $\Phi^0\to \gamma\gamma$ decay is of the same order of magnitude that $\Phi^P\to \gamma\gamma$. In contrast, $\mathrm{Br}(\Phi^0\to \gamma Z)\sim10\,\mathrm{Br}(\Phi^P\to \gamma Z)$, $\mathrm{Br}(\Phi^0\to WW)\sim10^2\,\mathrm{Br}(\Phi^P\to WW)$ and $\mathrm{Br}(\Phi^0\to ZZ)\sim10^8\,\mathrm{Br}(\Phi^P\to ZZ)$; the last discrepancy is mainly due to that the $\Phi^0\to ZZ$ decay is induced at tree level while the $\Phi^P\to ZZ$ decay appears at one-loop level.

Although our study does not represent a detailed calculation of the production mechanism of a new exotic scalar particle via gluon fusion at the LHC, it could offer a most complete experimental guide in order to search for new exotic scalar particles with masses of the order of TeVs~\cite{us}. Alternatively, the production mechanism of new heavy scalars could be studied in the context of the Compact Linear Collider (CLIC) bearing in mind that this collider would offer much cleaner collisions also to a high integrated luminosity~\cite{CLIC}.

\section{Conclusions}\label{CONCLU}
The LTHM is based on a nonlinear sigma model together with a $SU(5)$ global symmetry and a gauged subgroup $[SU(2)_1\otimes U(1)_1]\otimes[SU(2)_2\otimes U(1)_2]$, where the presence of new particles with masses at energy scales of TeV is proposed. Among all the variety of new particles predicted, a new neutral pseudoscalar particle, identified as $\Phi^P$, is the main subject of this work. In specific, to explore the predictability of the LTHM the $\Phi^P\to WW, VV, gg$ decays were studied. Even though the parameter space of the LTHM is severely restricted by electroweak precision data, there is still opportunity to test this model at low energies between 2-4 TeV. We have settled down an analysis region from 2 TeV to 4 TeV for the energy scale $f$, which indicates a mass interval for $m_{\Phi^P}$ between 1.66 TeV and 3.32 TeV, respectively. On this energy range, it has been found that the relevant branching ratios correspond to the $\Phi^P\to gg$ and $\Phi^P\to WW$ decays, being at most of the order of $10^{-4}$ and $10^{-6}$, respectively. A roughly estimate for the production cross section of the $\Phi^P$ boson via gluon fusion was implemented. For $f=2$ TeV, the numerical results tell us that in the last planned stage of operation of the LHC, events corresponding to the $\Phi^P\to WW$ decay would be very difficult to detect, however, the possibility of detection for the $\Phi^P\to gg$ process is more likely to be present.

\section*{Acknowledgments}
This work has been partially supported by CONACYT, SNI-CONACYT, and CIC-UMSNH.\\


\end{document}